\documentclass[rmp,aps,twocolumn]{revtex4}
\usepackage{graphicx}
\usepackage{amsmath,amssymb}

\newcommand{\tr}{\mathop{\rm Tr}}
\newcommand{\im}{\mathop{\rm Im}}
\newcommand{\re}{\mathop{\rm Re}}

\newcommand\ktl{K_{\rm TL}}
\newcommand\ks{K_{\rm SP}}
\newcommand\ym{x_{\rm M}}
\newcommand\ca{${\rm Ca}^{2+}$\ }
\newcommand\po{p_{\rm O}}
\newcommand\z{{\cal Z}}

\newcommand\I{{\rm i}}
\newcommand\B{{\cal B}}
\newcommand\F{{\cal F}}
\newcommand\ttl{T_{\rm TL}}
\newcommand\kb{K_{\rm B}}
\newcommand\db{b}
\newcommand\omc{\omega_{\rm C}}

\begin{document}
\title{Two adaptation processes in auditory hair cells
  together can  provide an active amplifier}
\author{Andrej Vilfan}
\altaffiliation[Present address: ]{J. Stefan Institute, Jamova 39, 1000
  Ljubljana, Slovenia}
\author{Thomas Duke}
\affiliation{Cavendish Laboratory, Madingley Road, Cambridge CB3 0HE, UK}
\email{av242@cam.ac.uk}
\date{15. May 2003}

\begin{abstract}
The hair cells of the vertebrate inner ear convert mechanical stimuli to
electrical signals.  Two adaptation mechanisms are known to modify the ionic
current flowing through the transduction channels of the hair bundles: a rapid
process involves \ca ions binding to the channels; and a slower adaptation is
associated with the movement of myosin motors.  We present a mathematical model
of the hair cell which demonstrates that the combination of these two
mechanisms can produce `self-tuned critical oscillations', i.e. maintain the
hair bundle at the threshold of an oscillatory instability. The characteristic
frequency depends on the geometry of the bundle and on the \ca dynamics, but is
independent of channel kinetics. Poised on the verge of vibrating, the hair
bundle acts as an active amplifier. However, if the hair cell is sufficiently
perturbed, other dynamical regimes can occur. These include slow relaxation
oscillations which resemble the hair bundle motion observed in some 
experimental preparations. 
\end{abstract}

\maketitle

\section*{Introduction}

Hair cells of the inner ear detect mechanical stimuli by deflections of
the hair bundle, which open tension-gated transduction channels in the
cell membrane to admit cations from the endolymph. The current is mostly
composed of potassium ions, but also includes a small quantity of
calcium ions \citep{Lumpkin.Hudspeth1997}. Experiments in which hair
bundles are held at a fixed deflection, using a microneedle, have
revealed two adaptation processes that modify the transduction current,
causing it to decline with time
\citep{Howard.Hudspeth1987,Howard.Hudspeth1988,Wu.Fettiplace1999,Holt.Corey2000,Holt.Gillespie2002}.
The first, which can occur on a sub-millisecond time scale, is dependent
on the concentration of \ca in the endolymph. It is believed to be a
consequence of \ca entering the hair bundle and binding to the
transduction channels, making them more likely to close. The second,
which takes up to hundreds of milliseconds, is also dependent on \ca
concentration, and is thought to be caused by the movement of 
adaptation motors attached to the transduction channels, which
adjusts the tension that gates them.

A number of experiments have demonstrated that hair bundles can generate
active oscillations, both in preparations of dissected tissue
\citep{Martin.Hudspeth2000,Benser.Hudspeth1993,Crawford.Fettiplace1985} and
{\it in vivo} \citep{Manley.Yates2001}.  It has been shown on theoretical
grounds that the active oscillatory mechanism can provide an explanation for
the high sensitivity and sharp frequency selectivity of auditory receptors
\citep{Choe.Hudspeth1998,Camalet.Prost2000,Eguiluz.Magnasco2000,Ospeck.Magnasco2001}
if the system is poised on the verge of spontaneous oscillation. This critical
point corresponds to a Hopf bifurcation of the active dynamical system.  It
has been suggested that a feedback mechanism operates to maintain the system
at the critical point where it is most sensitive, leading to the concept of
\emph{self-tuned critical oscillations}.  \citep{Camalet.Prost2000}.

The purpose of this article is to provide a coherent model for the
active amplifier in hair cells, based on the combined action of
the two known adaptation mechanisms. 
In our model the interaction of \ca ions with the transduction channels 
generates an oscillatory instability of the hair bundle. Working on a
slower time scale, the adaptation motors act to tune this dynamical system
to the critical point, at which it just begins to oscillate. The two
adaptation mechanisms together constitute a dynamical system
displaying a {\it self-tuned Hopf bifurcation}
\citep{Camalet.Prost2000}. Poised at the oscillatory instability, the
hair bundle is especially responsive to faint sounds.

Our model differs from two published models of hair-bundle oscillations. Choe,
Magnasco and Hudspeth (\citeyear{Choe.Hudspeth1998}) have previously proposed
that channel reclosure induced by \ca ions can generate a Hopf instability. In
their scheme the oscillation frequency is controlled by the kinetics of the
transduction channel. In our model the frequency of spontaneous oscillations
is governed, instead, by the architecture of the bundle and by the rate at
which the calcium concentration equilibrates.  This is more consistent with
the finding that the rate at which the transduction channel switches is faster
than the characteristic frequency, especially in low-frequency hair cells
\citep{Corey.Hudspeth1983b}.  \citet{Camalet.Prost2000} have suggested that
the Hopf instability is generated by motor proteins in the kinocilium, and
that regulation to the critical point is accomplished by a feedback mechanism
involving calcium. Their model is thus the converse of that presented here,
but shares the underlying feature of self-tuned criticality.  The current
model accords better with the observation that some hair bundles lack a
kinocilium and others do not change their behavior significantly when the
kinocilium is interfered with \cite{Martin.Hudspeth2003}.

In addition to demonstrating that the two adaptation processes can combine to
provide an active amplifier, we derive expressions that indicate how
properties like the characteristic frequency of critical oscillations, and the
sensitivity of response to sound waves, depend on parameters like the bundle
height and the number of stereocilia. We also investigate how the dynamics can
be modified if conditions are perturbed, and show that the bundle can exhibit
slow relaxation oscillations which differ in many aspects from critical
oscillations, but resemble the active movements recently observed in
experiments on hair cells from the frog sacculus \citep{Martin.Hudspeth1999}.

\section*{Model}

\subsection*{Hair bundle}
\begin{figure}[htbp]
    \begin{flushleft}
(a) \includegraphics{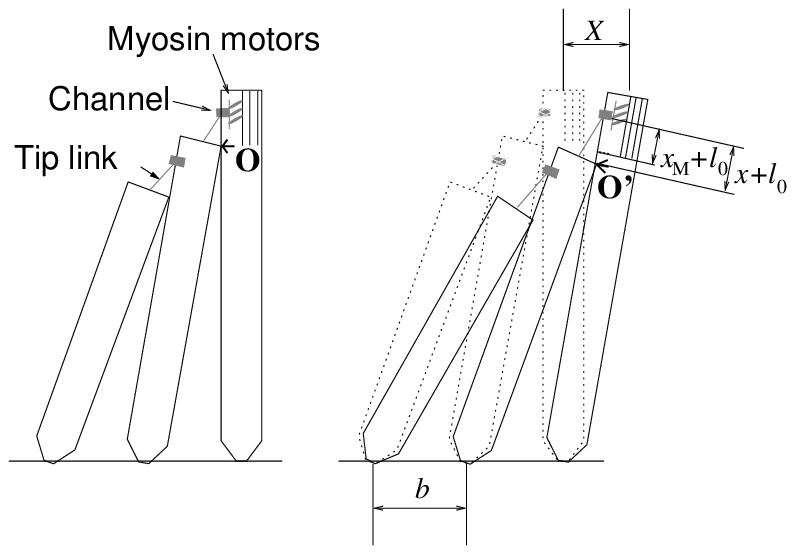}\\
        (b)\includegraphics{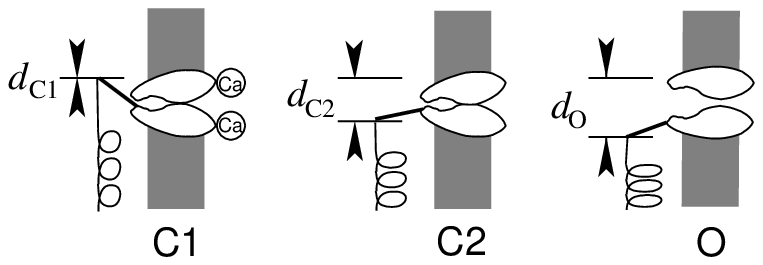}\\
    \end{flushleft}
  \caption{(a) Schematic representation of a hair bundle.
  (b) Three-state model of the transduction channel.
Note that \ca binds to the channel when it is already in a
closed state. The presence of \ca therefore does not directly cause channel
re-closure, but it does stabilize the closed state and delay re-opening.}
  \label{fig_bundle}
\end{figure}

Hair bundles in different species, or from hair cells with different
characteristic frequencies within the same organ of thte internal ear of a
particular animal, can vary in shape and size, and also in the way that they
are coupled to external mechanical vibrations \citep{Hudspeth1997}.  Generally,
one can distinguish between free-standing hair bundles (e.g. some reptiles,
mammalian inner hair cells) and those who tips are embedded in an accessory
structure (e.g. other reptiles, birds, amphibial saccular hair cells). In
experiments on dissected tissue, the accessory structure is usually removed.
In this paper we will restrict our discussion to the simplest situation of a
free-standing hair bundle.  A bundle that is directly connected to the
tectorial membrane could be described in a similar fashion, but using modified
expressions for the stiffness, the friction and the coupling to the external
signal.

The hair bundle (see Fig.~\ref{fig_bundle}a) is composed of $N$ stereocilia
which slope up against each other, the longest of which has length
$L$. Typically, $N$ and $L$ are inversely related, with the range $N=20-300$
\citep{Holt.Corey2000} corresponding to $L = 30-2 \,{\rm \mu m}$. Each
stereocilium is joined to the next by an elastic filament -- the tip link --
which connects to a tension-gated transduction channel in the cell membrane
(\citeauthor{Corey.Hudspeth1983b}, \citeyear{Corey.Hudspeth1983b}; for a
review see \citeauthor{Markin.Hudspeth1995b},
\citeyear{Markin.Hudspeth1995b}). The tension in the tip link can be adjusted
by a group of myosin motors, which are connected to the transduction channel
and move it along the actin filaments inside the stereocilium
\citep{Hudspeth.Gillespie1994}.

To set up a coordinate system, we consider first a free-standing bundle.
Within a stereocilium, we label the location at which the stereocilium touches
its shorter neighbor {\bf O}. The position of the channel (and the attached
motors) relative to {\bf O} is denoted by $l_0+\ym$, where $l_0$ is the length
of a tip link at rest and $\ym$ its extension under tension. When the tip of
the bundle is deflected through a lateral displacement $ X$, each stereocilium
pivots at its base and adjacent stereocilia are sheared by a displacement
$\gamma X$. The geometric factor $\gamma$ may be written as $\gamma\approx
\db/L$, where $\db \approx 1 \, \mu{\rm m}$ is the distance between the roots
of neighboring stereocilia \citep{Howard.Hudspeth1988a,Howard.Hudspeth1988}. We
label the new location at which the stereocilium touches its shorter neighbor
{\bf O'}.  The position of the channel relative to {\bf O'} is denoted by $l_0+x$.
It follows that
\begin{equation}
\label{eq_geom}
 \gamma X = x-\ym \;.
\end{equation}

The movement of the bundle is countered by an elastic restoring force,
which has two contributions. First, the deformation of the stereocilia at
their base provides an effective bundle stiffness $\ks = \kappa
N/L^2$, where $\kappa$ is the pivotal stiffness of a stereocilium.
 The measured value $\kappa \approx 1.5 \times 10^{-16} \,{\rm Nm/rad}$
\citep{Crawford.Fettiplace1985,Hudspeth.Martin2000} indicates that $\ks$ varies from
$0.01\,{\rm pN/nm}$ for tall bundles to $10\,{\rm pN/nm}$ for short
ones. The second contribution, which scales in the same way with bundle
size and is of similar magnitude, comes from the tension $\ttl$ in the tip
links. Movement of the bundle is also resisted by the viscous drag of the
surrounding fluid  which
leads to a friction coefficient $\zeta \sim \eta L$ for tall bundles, where
$\eta$ is the  
viscosity of the endolymph.
Thus, when an external force $F$ is applied
to the tip of the bundle, its equation of motion is
\begin{equation}
\label{eq_friction}
\zeta \dot X = - N \gamma \ttl(x, C) - \ks X\ + F\;,
\end{equation}
 (throughout the paper we use the convention $\dot X(t) \equiv
\frac{dX}{dt}$). 
Here, we have written $\ttl = \ttl(x, C)$, anticipating
that the tension in the tip links depends on the local intracellular
\ca concentration $C$, as well as varying with the position
$x$. For future reference, we note that the mechanical relaxation time
of the hair bundle, $\tau_{\rm mech} = \zeta/\ks$, varies from a few
microseconds to tens of milliseconds.

\subsection*{Transduction channel}

The mechanically-gated transduction channel is the central component of the
model, since we postulate that hair bundle oscillations are driven by the force
that is generated as the channels change conformation stochastically, as first
suggested by \citet{Howard.Hudspeth1988}. It is generally assumed that
transmembrane channel proteins can exist in a number of discrete states, some
of which are closed and some open. As indicated in Fig.~\ref{fig_bundle}b, we
additionally suppose that the channel protein incorporates a lever arm which
amplifies the small structural changes that occur when the channel switches
state, and thereby appreciably modifies the tension in the adjoining tip link 
 \citep{Martin.Hudspeth2000}.
If $p_i$ is the probability that the channel is in state $i$, and $d_i$ is the
position of the lever arm in that state, then the average tension may be
written $ \ttl =\sum_i \ktl (x-d_i) p_i$. Here, $\ktl$ is the elastic constant
of the tip link (or possibly that of a more compliant element in series with it
\citep{Kachar.Gillespie2000}), whose value obtained from micromechanical
measurements is
$\ktl \approx 0.5\,{\rm pN/nm}$ \citep{Martin.Hudspeth2000}. In the steady
state the probabilities $p_i=w_i/\left(\sum_j w_j\right)$ are given by the Boltzmann factors $w_i =
\exp \left[- (\frac{1}{2} \ktl (x - d_i)^2 -G^0_i)/k_BT \right]$, where $G^0_i$ is
the free energy of state $i$.  Measurements of the transduction current as a
function of bundle displacements in the frog's sacculus are well
fitted with a three-state model of the channel \citep{Corey.Hudspeth1983b},
with two closed states ($i=\rm{C1,C2}$) and one open state ($i={\rm O}$). The
fit indicates that the lever-arm movement is about 5 times greater for the
transition $\rm{C1 \rightarrow C2}$ than for the transition $\rm{C2 \rightarrow
  O}$, while more recent data \citep{Martin.Hudspeth2000} suggests that the
total movement is about 8\,nm. We therefore assign the lever arm displacements
$d_{\rm C1}=0\,{\rm nm}$, $d_{\rm C2}=7\,{\rm nm}$ and $d_{\rm O}=8.5\,{\rm
  nm}$. The free energies $G^0_i$ depend on the local \ca concentration $C$
within the stereocilia \citep{Howard.Hudspeth1988}, which is strongly
influenced by the entry of \ca ions through the channel
 \citep{Lumpkin.Hudspeth1995}. A higher \ca concentration favors closure of the
channel, thereby providing a negative feedback. We model this dependence
empirically, according to experimental observations
\citep{Corey.Hudspeth1983b}, as an effective shift in the channel gating point
proportional to the logarithm of the \ca concentration
\begin{equation}
\label{eq_po}
p_i (x,C)=p_i \left(x-x_s(C)\right) \;, \qquad
x_s(C)=D \ln \frac{C}{C_0} \;,
\end{equation}
where the constant $D$ has an empirical value of about $4\,{\rm nm}$ in the
bullfrog's sacculus \citep{Corey.Hudspeth1983b} and $C_0$ is the \ca
concentration in the free-standing bundle. One possible interpretation of this
functional form, indicated in Fig.~\ref{fig_bundle}b, is that \ca binds to
state C2, causing it to switch to state C1 and thereby stabilizing the closed
state.  We emphasize that our model does not rely on the channel having three
states.  A two-state channel (eliminating state C2) would give substantially
similar results.

An important aspect of our model is that we assume that the channel kinetics
are rapid compared to the bundle oscillations and the calcium dynamics.  The
evidence for this assumption comes, for instance, from saccular hair cells
that detect frequencies up to $150\,{\rm Hz}$ and  
have channel opening times around $100\,{\rm \mu s}$
\citep{Corey.Hudspeth1983b}.  We can therefore consider the different channel
states to be thermally equilibrated with one another, and chemically
equilibrated with the \ca ions. Then we can split the tip link tension $\ttl$
into two components: $\ktl x$, the passive elastic force that the tip link
would contribute in the absence of the transduction channel; and $f$, the
active contribution arising from the channel switching states, also called
\emph{gating force}:
\begin{equation}
\label{eq_force}
\ttl(x,C)=\ktl x + f\left(x-x_s(C)  \right) \;, 
\end{equation}
where $f\left(x,C \right)= -\sum_i \ktl d_i \,p_i\left(x,C \right)$. A
relationship between the gating force and the channel opening probability,
which holds in any model with a single open state, reads $f(x,C)\equiv k_BT
\frac {\partial \left(\ln p_{\rm O} (x,C) \right)}{\partial x} -\ktl d_{\rm O}$
\citep{van.Netten.Kros2000}. 

\subsection*{Calcium dynamics}

The dynamics of the calcium concentration in a stereocilium is influenced
by a number of processes. As well as the influx of \ca through the
transduction channel, there is an outflux due to calcium pumps. The ions
also diffuse freely, or in association with mobile buffers, and can
additionally bind to fixed buffers. A detailed
model of this dynamics was developed by \citet{Lumpkin.Hudspeth1998}, but
for our purposes it suffices to use a simplified linearized model in which
the \ca concentration on the inner side of the channel obeys the equation
\begin{equation}
\label{eq_dotc}
\dot C = -\lambda \left [ C - C_B - C_M \po (x, C) \right] \;.
\end{equation}
Here $C_B$ is the steady-state concentration which would occur if the channel
were blocked, originating from diffusion from other parts of the cell; $C_M$
is the maximal additional contribution from influx through an open
transduction channel, and is of order $10\,\mu {\rm M}$.  The relaxation rate
of calcium concentration fluctuations, $\lambda$, is an important parameter in
our model.  We expect the relaxation time $\tau_{\rm chem} = 1/\lambda$ to be
of the order of milliseconds, corresponding to the measured fast adaptation
time \citep{Ricci.Fettiplace1997}. Note that the slow \ca kinetics (along with
fast channel switching) is a key difference between our model and that of
\citet{Choe.Hudspeth1998}, where the converse situation is assumed. The slow
\ca dynamics might seem at odds with the fact that the binding sites on the
channel are reached by the incoming ions very quickly.  But in our model, the
channel is in state C2 when \ca binds, i.e. it is closed
(Fig.~\ref{fig_bundle}b). It therefore typically senses a concentration which
is determined by diffusion within a larger volume, and which also depends on
the interaction of \ca with fixed and mobile buffers.  Because a \ca ion binds
to one of the buffers in less than $1\,\mu{\rm s}$
\citep{Lumpkin.Hudspeth1998}, the concentration of free \ca equilibrates with
that bound to the buffers within a few microseconds of channel closure
O$\rightarrow$C2.  The time-scale of \ca relaxation is therefore determined by
buffering, diffusion to distant parts of the stereocilium and the activity of
\ca pumps. Indeed, the influence of buffers on the fast adaptation process has
been observed experimentally \citep{Ricci.Fettiplace1997}.

\subsection*{Adaptation motors}

Experiments in which a vertebrate hair bundle is suddenly displaced
\citep{Howard.Hudspeth1987,Wu.Fettiplace1999} show a slow adaptation of the
channel current, on the scale of tens of milliseconds.  The transduction
channel is attached to the actin cortex within a stereocilium by a number of
myosin 1C (formerly I$\beta$) motors.  There is strong evidence that these
molecular motors mediate slow adapatation
\citep{Assad.Corey1992,Gillespie.Corey1997,Holt.Gillespie2002}, presumably by
maintaining the proper tenstion in the tip link \citep{Hudspeth.Gillespie1994}.
For simplicity, we assume that the motors have a linear force-velocity
relation, so that
\begin{equation}
\label{eq_motfv}
\dot x_{\rm M} =v_M \left[ 1 - \frac{\ttl}{n_M F_M} \right] \;,
\end{equation}
where $n_M$ is the number of motors bound to the actin, $F_M$ is the stall
force of an individual motor and $v_M$ is its zero-load velocity. We take
values typical for a motor protein, $F_M \approx 1 \,{\rm pN}$ and $v_M \approx
0.3 \,{\rm \mu m/s}$ and suppose that there are $N_M \approx 20$ motors per
channel (experimental data suggest $N_M\lesssim 130$
\citep{Gillespie.Hudspeth1993}).

Calcium has been shown to influence the action of the adaptation motor 
\citep{Hacohen.Corey1989,Ricci.Fettiplace1998}, possibly by increasing its 
rate of detachment from actin.  The number of bound motors is
therefore not constant, but obeys the dynamical equation 
\begin{equation}
\label{eq_mot}
\dot n_M=\omega_M \left[-n_M g(C)+(N_M-n_M) \right] \;.
\end{equation}
where $\omega_M\approx 200 \,{\rm s^{-1}}$ is the myosin binding rate and
$g(C)$ is a function describing the dependence of the detachment rate on the
\ca concentration.

\section*{Results and Discussion}

\subsection*{Linear stability analysis}

In order to examine the dynamical stability of the hair bundle
in the absence of a driving force, $F=0$, we make use of the fact that the
adaptation motors operate on a slower time scale than both the mechanical
relaxation of the bundle and the relaxation of the calcium concentration.
We can therefore neglect their motion for now, and suppose that $\ym$ is
fixed. This leaves two coupled first-order differential equations of
motion: Eq.~\ref{eq_dotc} (supplemented with Eq.~\ref{eq_po}) for the
rate of change of \ca concentration; and by combining Eqs.\ 
\ref{eq_geom}, \ref{eq_friction}
\& \ref{eq_force}, the following equation for the motion of the
transduction channels:
\begin{equation}
\label{eq_ydot}
\dot x=-\frac{1}{\zeta}\lbrace N \gamma^2 \left [ \ktl
x + f\left(x-x_s(C) \right) \right]+\ks (x - \ym)\rbrace
\;. 
\end{equation}

The system has one or more fixed points, given by $\dot x=0$ and $\dot
C=0$, whose stability can be determined from the Jacobian matrix
\begin{equation}
{\bf J}= \left(\begin{array}{cc}\frac{\partial \dot x}{\partial x} & 
\frac{\partial \dot C}{\partial x}
\\
\frac{\partial \dot x}{\partial C}
&
\frac{\partial \dot C}{\partial C}
\end{array}\right) \;,
\label{eq_jacobi}
\end{equation}
which has trace and determinant
\begin{eqnarray}
\label{eq_trace}
\tr {\bf J} &=& -\frac 1 \zeta \left[  \kb + N \gamma^2 f'
\right] -\lambda
\left( 1+ \mu \right) \;, \\
\label{eq_determinant}
\det {\bf J} &=& \frac {\lambda}{\zeta} \left[ \kb 
\left( 1+  \mu \right) + N \gamma^2 f' \right] \; ,
\end{eqnarray}
(throughout the paper we use the convention $f'(x) \equiv
\frac{\partial f}{\partial x}$, $p_{\rm O}'(x) \equiv \frac{\partial p_{\rm
    O}}{\partial x}$  and $x_s' \equiv \frac{\partial x_s}{\partial  C}$).
Here, $\kb = \ks + N \gamma^2 \ktl$ is the combined passive elasticity
of the bundle due to the stereocilia and the tip links; and 
$\mu = C_M x_s' \po'$ is a dimensionless combination of variables which
has a value close to unity for the choice of channel parameters specified
above (for $C_B=0$ it can also be expressed as $\mu = D (f+\ktl d_{\rm
  O})/k_BT$). 
A fixed
point is stable if both eigenvalues of ${\bf J}$, evaluated at the fixed
point, have a negative real part. This is equivalent \citep{Strogatz1994}
to the two conditions:
$\det {\bf J} >0$, and $\tr {\bf J} <0$.
The number of fixed points is most readily appreciated by considering the
form of Eq.~\ref{eq_ydot} for the velocity $\dot x$ as a
function of displacement $x$ (see Fig.~\ref{fig_fy}). Two distinct
situations can arise depending on the strength of the calcium feedback:

\begin{figure}
    \begin{center}
      \includegraphics{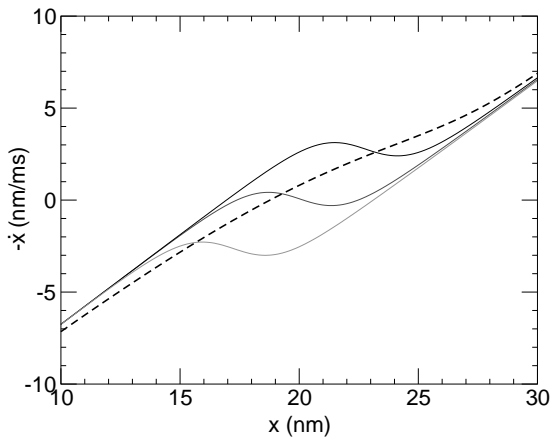}
    \end{center}
  \caption{
    Velocity $\dot x$ as a function of displacement 
    $x$ for a fixed location of the motors $\ym$. If the \ca concentration
    is held constant, $\dot x(x,C={\rm const})$ typically has a
    region of negative slope (solid lines, for three different 
    hypothetical values of
    $C$). For a certain range of values of $C$, the system can then 
    have three fixed points. But if the \ca concentration is allowed
    to adjust, the curve $\dot x(x,C=C(x))$, where $C(x)$ is the
    steady-state solution of Eq.~\ref{eq_dotc}, rises monotonically 
    with $x$ (dotted line). There is then a single fixed point.  The channel
    parameters are $G^0_{\rm C1}=0$, $G^0_{\rm C2}= 60 \,{\rm zJ}$,
    $G^0_{\rm O}= 70 \,{\rm zJ}$,
    $d_{\rm C1}=0\,{\rm nm}$, $d_{\rm C2}=7\,{\rm nm}$, 
    $d_{\rm O}=8.5\,{\rm nm}$ and $D=4\,{\rm nm}$. With these channel
    parameters $\ktl d_{\rm O}^2 \approx 9 k_B T$ (in a simplified two-state
 channel model neglecting the stereociliary stiffness $\ks$ the condition for the occurrence of a negative slope is
    $\ktl d_{\rm O}^2 > 4 k_B T$).}
  \label{fig_fy}
\end{figure}

\begin{figure}[tb!]
  \begin{tabular}{lr}
  (a)&\includegraphics{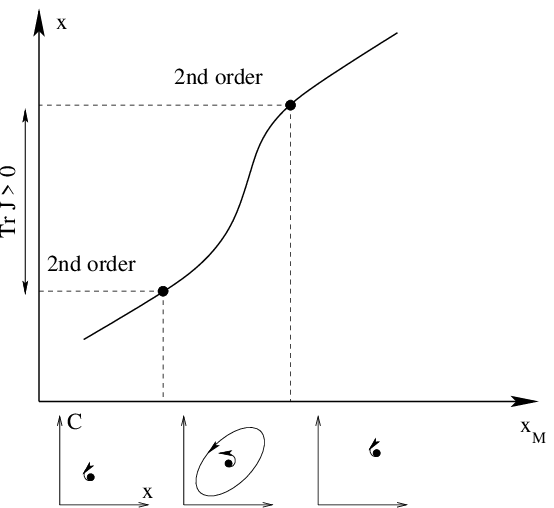}\\
  (b)&\includegraphics{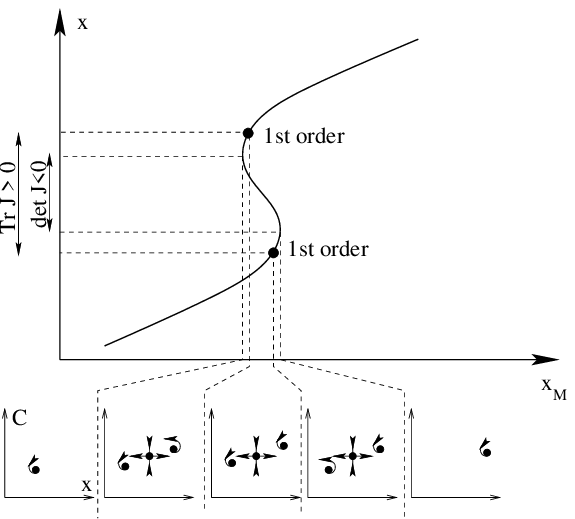}
  \end{tabular}
\caption{
  Fixed points of the displacement $x$ as a function of the location
  of the adaptation motors $\ym$. Trajectories in the plane $(x, C)$
  are indicated schematically below. (a) The situation with a weak \ca
  feedback. The system displays a variety of dynamical regimes,
  including a region of bistability. (b) With strong \ca feedback
  there is a single fixed point which becomes unstable (and encircled
  by a stable limit cycle) for an intermediate range of values of
  $\ym$. We expect the hair cells to use this regime {\it in vivo}.}
\label{fig_fixed}
\end{figure}

\begin{figure}
   \begin{center}
      \includegraphics{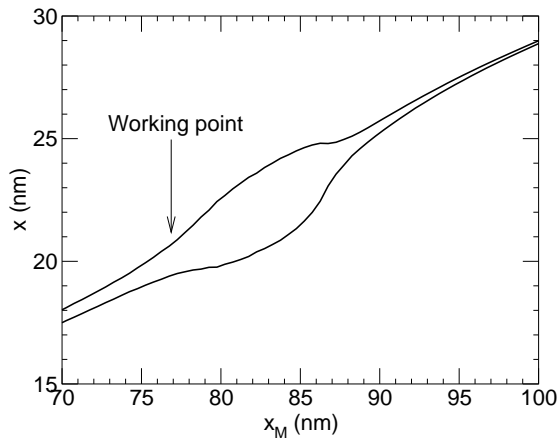}
    \end{center}
  \caption{
    Displacement $x$ as a function of the location of
    the adaptation motors $\ym$, for the proposed {\it in vivo} situation
    corresponding to Fig.~\ref{fig_fixed}b. Between the two bifurcation
    points, $x$ oscillates between a minimum and a maximum value
    shown by the two curves. The motor-mediated self-tuning mechanism
    (see text) maintains the system at a working point
    close to the first Hopf bifurcation.  This result was obtained
 from a stochastic simulation with full channel kinetics, but
 in
    the absence of external (thermal) noise. The upper and 
    lower line show the average maximum and minimum values of the displacement
    during one period of oscillation.  Note that even in the quiescent phase the
    amplitude is non-zero, due to channel noise.  The
    noisy motion in the quiescent phase has a very different
    spectrum from the oscillating phase, however. }
  \label{fig_yym}
\end{figure}

(i) Consider first the case of weak feedback, so that the \ca
concentration $C$ in the stereocilia is approximately constant.  Typically
$\dot x(x, C=\rm{const})$ varies {\em non-monotonically} with $x$. This is a
consequence of a region of effective {\em negative elasticity}, caused by the
redistribution of the transduction channel states when the bundle is moved
 \citep{Howard.Hudspeth1987,Martin.Hudspeth2000}.  At large displacements --
both positive and negative -- the tension $\ttl$ in the tip links is determined
by their passive elasticity. But at intermediate displacements, the occupancy
of the closed state C1 declines rapidly with increasing $x$; the resulting
force $f$, caused by the swing of the lever arm, is negative and {\em reduces}
the tension in the tip links so that $\ttl$ decreases with increasing $x$. 
This
effect has been termed `gating compliance' by \citet{Howard.Hudspeth1987}. As
indicated in Fig.\,\ref{fig_fy}, there will then be either one or three fixed
points. According to the index theorem \citep[][Sect.~6.8]{Strogatz1994} out of
$2n+1$ fixed points, $n$ will be 
saddle-points.  Therefore, if the system has three fixed points, the central
fixed point is a saddle point and the other two may be either stable
or unstable. The number of fixed points depends on
the values of the channel parameters, but also on the motor position $\ym$.
Their stability depends additionally on the values of the dynamical model
parameters ($\zeta$ and $\lambda$). A typical example of the location and
nature of the fixed points, as a function of the motor position $\ym$, is shown
in Fig.\,\ref{fig_fixed}a; in this particular case, the bundle displays a
region of {\em bistability}. We shall return to this situation later, to
discuss the dynamical behavior of the hair bundle in recent {\em in vitro}
experiments. 

(ii) When the calcium feedback is operational, to determine the fixed points
it is appropriate to consider the curve $\dot x (x, C=C(x))$, where $C(x)$ is
the steady-state \ca concentration at fixed $x$, given by $\dot C = 0$ in
Eq.~\ref{eq_dotc}.  If the feedback is sufficiently strong, $\dot x$ varies
{\em monotonically} with $x$ (Fig.~\ref{fig_fy}) ; the readjustment of the
channel states caused by the binding of \ca ions provides a positive
contribution to the active force $f$, eliminating the region of negative
elasticity. The condition for this to occur is that $\det {\bf J} > 0$ along
the curve $C(x)$, or
\begin{equation}
\mu > -\frac{N \gamma^2}{\kb} f'-1\;.
\end{equation} 
In this case, there is a single fixed point (Fig.~\ref{fig_fixed}b), which may
be either stable (if ${\tr{\bf J}} < 0$) or unstable (if ${\tr{\bf J}} >
0$). In the latter situation, the bundle undergoes limit-cycle
oscillations. Significantly, the stability depends on the location of the
motors {$\ym$}, in addition to the values of the other model parameters.
Typically the critical point at which ${\tr{\bf J}}=0$ is a
\emph{supercritical} Hopf bifurcation, which means that the system is
continuously controllable, passing smoothly from the quiescent to the
vibrating state as indicated in Fig.\,\ref{fig_yym}. Thus by moving along the
stereocilia, the motors can determine whether the bundle stays still, or with
what amplitude it oscillates spontaneously.

\subsection*{Frequency of limit-cycle oscillations}

We propose that the regime of strong calcium feedback, in which the bundle can
oscillate spontaneously, is the regime that is relevant for auditory hair
cells in normal physiological condititions.  The control parameter that
determines the stability of the bundle is $\epsilon=\frac 12 \tr {\bf J}$ and
the Hopf bifurcation occurs at $\epsilon=0$. It is worth emphasizing that
within the scope of a non-inertial model with two variables ($x$ and $C$), in
order for the bundle to oscillate spontaneously, it is essential that either
the bundle displays a region of negative elasticity at constant \ca
concentration, or that there is a positive feedback in the \ca dynamics.  The
second option would imply that an increase in the intracellular \ca
concentration would enhance the \ca inflow, which contradicts the present
experimental evidence.  Our model only allows for the first option and from
Eq.~\ref{eq_trace} it can be seen that the critical condition $\epsilon=0$
requires that $K_B + N \gamma^2 f' < 0$, i.e. the passive bundle stiffness
must be more than compensated by the gating compliance.  Negative elasticity
has been measured in hair cells of the bullfrog sacculus
\citep{Martin.Hudspeth2000}, and has been interpreted as being due to gating
compliance \citep{Howard.Hudspeth1987} as discussed above.

For small, positive values of $\epsilon$, the bundle executes 
small-amplitude limit-cycle oscillations of frequency $\omc = \sqrt{\det {\bf
J}}$. From Eq.~\ref{eq_determinant} it follows that the
characteristic frequency of the hair bundle scales as
\begin{equation}
  \label{eq_scale_omega}
  \omc \sim \sqrt{ \lambda \frac {\kb}{\zeta} }  
  \quad \propto \sqrt{\lambda \frac{N}{L^3} }
\end{equation}
Note that the period of oscillation $T_{\rm C}$ is approximately given by the
geometric mean of the two relaxation times, $T_{\rm C} \sim \sqrt {\tau_{\rm
    mech} \tau_{\rm chem}}$. Little is known about the variability of the
calcium relaxation rate $\lambda$. If it is constant from cell to cell, the
characteristic frequency is determined only by the bundle geometry.  For the
range of bundle sizes found e.g. in the chick's cochlea (L = $1.5-5.5 \,{\rm
  \mu m}$, $N=300-50$ \citep{Tilney.Saunders1983}), the frequency of the
shortest bundles would then be approximately 17 times greater than the
frequency of the tallest bundles.  This is somewhat lower than the measured
frequency range (about 50\,Hz to 5\,kHz \citep{Manley.Gleich1991}). We note,
however, that to ensure the existence of an oscillatory instability for all
hair cells, $\epsilon=0$ from Eq.~\ref{eq_trace} must be fulfilled for each
cell. In addition, a strong mismatch between $\tau_{\rm mech}$ and $\tau_{\rm
  chem}$ would seriously limit the range of amplitudes for which active
amplification can be maintained.  This suggests that $\lambda$ may vary with
the geometry of the cell; ideally it would scale as $\lambda \sim \kb/\zeta$,
i.e. ${\tau_{\rm chem} \sim \tau_{\rm mech}}$. Such variation could be achieved
by a differential level of expression of calcium pumps or different buffer
concentrations.  In this case, the range would be enhanced to cover frequencies
differing by a factor of 200. In what follows, we shall assume that this
scaling (which is equivalent to $\lambda \sim \omega_{\rm C}$) holds.  Note
that in contrast to the model proposed by Choe, Magnasco and Hudspeth
(\citeyear{Choe.Hudspeth1998}) no variation of transduction channel kinetics is
necessary in order to achieve a broad range of characteristic frequencies.
Instead a simultaneous variation of the hair bundle geometry and the \ca
relaxation rate is required.

\subsection*{Critical oscillations} 

It has been proposed on general theoretical grounds that the frequency
selectivity of hair bundles, as well as their sensitivity to weak
signals, is conferred by their proximity to a dynamical instability
\citep{Camalet.Prost2000,Eguiluz.Magnasco2000}.
Poised on the verge of oscillating at a characteristic frequency
$\omc$, a hair bundle is especially responsive to periodic
disturbances at that frequency. To ensure robust operation, a control
system is required which maintains the hair bundle close to the critical
point at which it just begins to vibrate. The
general concept of a self-tuned Hopf bifurcation was therefore
introduced \citep{Camalet.Prost2000}, whereby an appropriate feedback
mechanism automatically tunes the system to the vicinity of the critical
point. For the moment, we will assume that
this mechanism  works effectively, so that each hair cell is poised at a
working point very close to its critical point, as indicated in
Fig.\,\ref{fig_yym}. 

In the vicinity of the bifurcation
the limit cycle can be described by expanding the
equations of motion about the fixed point as a power series, keeping terms
up to third order.  It is convenient to introduce a complex variable
\begin{equation}
\z=\Delta x + \alpha \Delta C,
\label{eq_z}
\end{equation}
 where $\Delta x$ and $\Delta C$ measure
the difference of $x$ and $C$ from their respective values at the
critical point. With the choice
\begin{equation}
\label{eq_alpha}
\alpha=\frac{J_{22}-\epsilon + \I \omc}{J_{12}} \;,
\end{equation}
where $J_{12}=\lambda C_M \po'$ and $J_{22}=-\lambda(1+\mu)$
(Eq.~\ref{eq_jacobi}), the equations of motion can be written as
\begin{equation}
\label{eq_complex}
\dot \z =(\epsilon + \I \omc) \z + {\cal Q} + \B |\z|^2 \z +
{\cal C} \;,
\end{equation}
where ${\cal Q}$ represents all quadratic terms and ${\cal C}$ all remaining
cubic terms.  The expression for the coefficient $\B$ is given in the
Appendix; for typical parameters, $\re\B$ is negative, in which case the
bifurcation is supercritical and therefore continuously controllable. To
leading order, the amplitude of the variable $\z$ is then given by ${\cal
A}=\sqrt{-{\epsilon}/{\re\B}}$, while all other terms cause only perturbations
to the trajectory \citep{Wiggins1990}. The solution for spontaneous critical
oscillations therefore reads
\begin{equation}
\Delta x(t)=\frac{\im (\alpha \bar \z(t))}{\im \alpha} \;, \quad
\Delta C(t)=\frac{\im \z(t)}{\im \alpha} \;, \quad
\z(t)={\cal A} e^{\I \omc t }
\label{eq_solution}
\end{equation}
 and the oscillation amplitude of $\Delta x$ is
$\frac{\left|\alpha\right|}{\im \alpha} {\cal A}$. 

\subsection*{Critical response to sound waves}

Sound waves or other vibrations entering the inner ear cause flow in the fluid
surrounding the hair bundles. If the (unperturbed) fluid motion at the height
of the bundle tip has amplitude $A$ and frequency $\omega$, the force exerted
on the bundle tip in Eq.~\ref{eq_friction} is $F=A \omega {\zeta_{\rm flow}}
\cos(\omega t)$. Here, $\zeta_{\rm flow} \sim \eta L$ is the friction
coefficient of a stationary bundle in a flowing fluid. Note that $\zeta_{\rm
flow}$ is not precisely equal to $\zeta$, which is the friction coefficient of
the bundle moving in a stationary flow, because the latter also includes a
contribution from shear flow between the stereocilia.  This contribution
scales as $\sim \eta L \times N\gamma^2$ and is negligible for tall bundles
but might become significant for short thick bundles. To preserve generality
we write $\zeta_{\rm flow}/\zeta = \nu$, where $\nu$ is a constant of order
unity which might vary with bundle geometry.
 
The external force leads to an additional driving term on the right hand
side of Eq.~\ref{eq_complex}, given by
\begin{equation}
\label{eq_F}
\F=\frac12 A \omega \gamma \nu e^{\I \omega t} \;.
\end{equation}

For sound frequencies close to the characteristic frequency of the bundle,
$\omega \approx \omc$, and amplitudes larger than the thermal noise, 
the response of the system is
 
\begin{equation}
\label{eq_gain_high}
\z=\sqrt[3]{\frac{|\F |}{-\re \B}} \;e^{\I \omega t}\;,
\end{equation}
 
where we have assumed that the bundle is 
poised precisely at the Hopf bifurcation, $\epsilon=0$.
 
Using the estimate $\re \B \sim -\omc/d_{\rm O}^{2}$ (see
Appendix) we obtain
\begin{equation}
\label{eq_gain_high_scale}
\Delta x \sim \sqrt[3]{\gamma \nu A  d_{\rm O}^2 } = \sqrt[3]{\nu
A d_{\rm O}^2
 \db /L}\;.
\end{equation}
This is the strongly compressive response characteristic of a Hopf
resonance \citep{Camalet.Prost2000,Eguiluz.Magnasco2000}.
Displacements of the transduction channel which produce a resolvable 
response, $\Delta x = 1-10\;{\rm nm}$, typically correspond to wave
amplitudes
$A = 1-1000\;{\rm nm}$ in the fluid.

For weak stimuli, it is important to take the effects of noise into
consideration. The buffeting of the hair-bundle by the Brownian motion in the
surrounding fluid will render the spontaneous critical oscillations noisy.  In
this regime, signals that yield a response that is smaller than the
spontaneous motion can nevertheless be detected, owing to phase-locking of the
noisy oscillations \citep{Camalet.Prost2000}.  The ability to detect phase
locking in neural signals despite no increase in the firing rate has been
experimentally confirmed in a number of species \citep{Narins.Wagner1989}.  In
the Appendix we provide an argument for the threshold of detection which
suggests that for high frequency bundles containing many stereocilia, the
smallest wave amplitude that can be detected may be as small as $A \approx
0.1\;{\rm nm}$.

\subsection*{Self-tuning mechanism}

In order to achieve the combination of frequency selectivity,  
amplification of weak signals and wide dynamic range, the system has
to be tuned to the close proximity of the Hopf bifurcation 
($\tr {\bf J}=0$ in Eq.~\ref{eq_trace}). We propose
that this is accomplished by a feedback control mechanism which
is mediated by the adaptation motors. The basic requirement is that when
the hair bundle is still, the force exerted by the motors
gradually increases so that they advance up the stereocilia
(increasing $\ym$), pushing the system into the unstable regime (see
Fig.~\ref{fig_yym}); but when the bundle is oscillating, the motor force
gradually declines, so that the motors slip back down the stereocilia,
returning the system to the quiescent state. It would appear that some type of
high-pass or band-pass filter, which can detect whether or not the bundle
is oscillating, must form an integral part of the feedback mechanism that
controls the motors.

It has long been known that the hair cells of some vertebrates have, in their
cell membranes, a system of potassium channels (BK) and voltage-gated calcium
channels that display a resonant response to oscillating input currents
\citep{Hudspeth.Lewis1988a}. It has been suggested that their purpose is to
sharpen the tuning of the signal passed to the auditory nerve
\citep{Crawford.Fettiplace1980}. It has alternatively been postulated that
this system may itself generate self-sustained Hopf oscillations
\citep{Ospeck.Magnasco2001} and could be the basis of active bundle
movement. We put forward a further possibility: One role of this system of
channels is to detect bundle oscillations and to control the adaptation motors
via a flux of calcium ions into the hair bundle (in fact, it is more likely
that the motors are regulated by \ca-controlled secondary messengers diffusing
into the hair bundle, rather than by \ca directly, but this would not
qualitatively alter the following argument).

We model the filtering characteristic of the electrical oscillator by a Lorentzian transmission function
\begin{equation}
  \label{eq_filter}
  V(\omega)=V_0+I_{\rm input}(\omega) R_{\rm F}
  \frac{\omega_{\rm F}^2}{\omega_{\rm F}^2-\omega^2+\I \omega 
  \omega_{\rm F}/ Q}
\end{equation}
where $V$ is the receptor potential, $R_{\rm F}$ is a channel resistance and
$I_{\rm input} \propto p_{\rm O}$ is the transduction current. The
mechanism operates best if the filter frequency $\omega_{\rm F}$  equals the
characteristic frequency $\omc$ of hair-bundle oscillations, but this matching
has to be encoded genetically or achieved by some mechanism beyond the scope of
our model. With the value
$Q\approx 10$ measured experimentally in the bullfrog's sacculus
\citep{Hudspeth.Lewis1988a}, the filter
distinguishes sufficiently well between the oscillating component of the
transduction current and the underlying background level.  The voltage-gated
calcium channels then provide a \ca flux varying (for example) exponentially
with the membrane potential, which in turn determines the offset concentration
in the stereocilia $C_B$, which appears in Eq.~\ref{eq_dotc}. We model it with
the following differential equation
\begin{equation}
  \label{eq_ca_simulation}
  \dot C_B= I_{\rm Ca} \frac1 {1+\exp(-q(V-V_{\rm Ca})/k_BT)} -\lambda_{\rm B} C_B\;,
\end{equation}
where the parameters $q$ and $V_{\rm Ca}$ describe the voltage sensitivity of
the ${\rm K}_{\rm Ca}$ channels, $I_{\rm Ca}$ is a measure of these channels'
contribution to the \ca flux in the region close to the transduction channel,
and $\lambda_{\rm B}$ is the relaxation rate of the \ca current.  For the
operation of tuning mechanism it is essential that $\lambda_{\rm B} \ll
\omega_C$. The contribution of $C_B$ is insignificant when the bundle is
quiescent, but grows rapidly if the bundle starts to oscillate.

The \ca flowing into the bundle from the cell body regulates the
oscillations in two ways. First, it reduces the control parameter
of the bifurcation directly, Eq.~\ref{eq_trace}. Second, \ca ions
binding to the myosin motors enhance their rate of detachment, via the
function $g(C)$ in Eq.~\ref{eq_mot}, thereby regulating the total force
that the team of motors produces. Here we take $g(C) \propto C^3$; the
precise functional form is not very important.

\begin{figure}
  \begin{center}
  \includegraphics{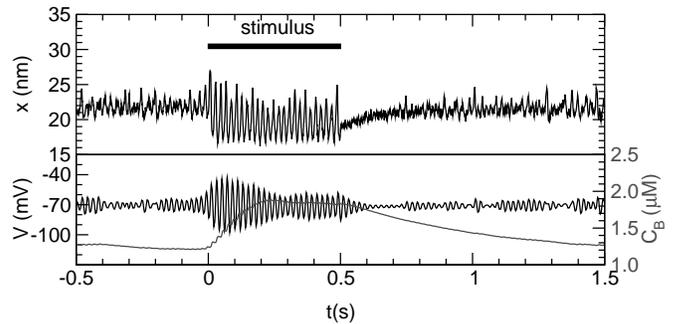}
  \end{center}
  \caption{Self-tuned critical oscillations.  In the absence of a stimulus, the 
    bundle performs noisy oscillations with a small amplitude.  A strong
    external stimulus elicits a response that can detune the system for some
    time.  Parameters values: $N=50$, $\ktl=0.5\,{\rm pN/nm}$, $\ks=0.15\,{\rm
      pN/nm}$, $\zeta=\zeta_{\rm flow}=0.65\,{\rm pN\,ms/nm}$, $\gamma=0.14$ and
    $A=40\,{\rm nm}$ ( the values of other parameters are listed in the
    last section of the Appendix ).}
  \label{fig_tuning}
\end{figure}

As shown in Fig.~\ref{fig_tuning}, this feedback control mechanism
can successfully tune the system to the immediate vicinity
of the Hopf bifurcation. In the absence of a sound stimulus, a bundle
executes noisy, self-tuned critical oscillations of low amplitude. If a
strong stimulus is then applied, the bundle displays the characteristic
response given by Eq.~\ref{eq_gain_high_scale}. The significant flow of
\ca into the cell elicited by this response detunes the system, so that
when the stimulus is cut, the bundle falls still. Gradually, however, the
motor-mediated self-tuning mechanism returns the system to its working
point, where critical oscillations just set in, and where the hair
bundle's sensitivity is maximal.

The hypothesis that the self-tuning mechanism is based on the detection of
oscillations in the transduction current could be tested experimentally by
disabling the electrical  oscillations in the cell body.  This could be done by
voltage-clamping the cell or by specifically blocking, for example, the
voltage-gated calcium
channels.  This would disable the self-tuning mechanism and therefore augment
the bundle oscillations.  On the other hand, if the electrical resonance is
involved in the generation of oscillations, disabling it should eliminate the
spontaneous bundle activity.

\subsection*{Slow relaxation oscillations}

\begin{figure}
   \begin{tabular}{cc}
      (a)&\includegraphics{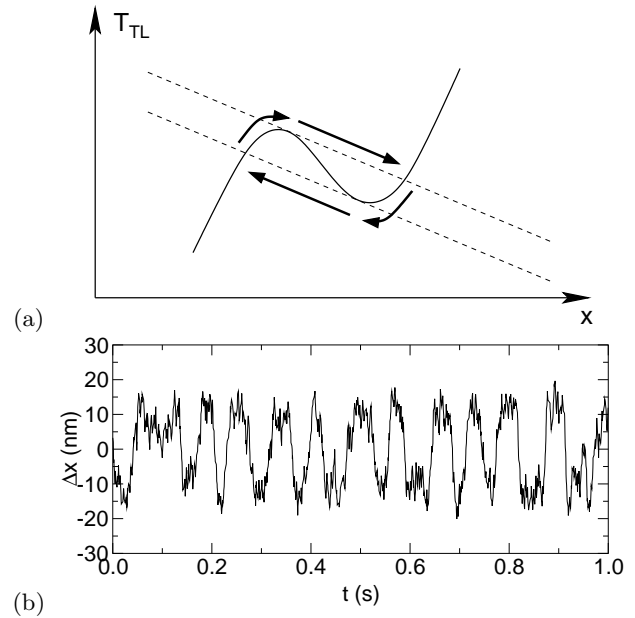}\\
      (b)&\includegraphics{fig6b}\\
    \end{tabular}
  \caption{Slow relaxation oscillator regime. (a) The solid line
    shows the tension-displacement relation for a channel, the dashed lines the
    contribution to tension from the passive bundle stiffness (which depends on
    the position of adaptation motors).  In a quasistationary state these two
    must be equal.  For a given position of the adaptation motors, there can
    be two stable solutions. The motors then move the system from one
    fixed point to the other.  (b) Relaxation oscillations have a lower
    frequency than and a different shape from the critical
    oscillations.  To achieve this regime the \ca concentration outside the
    bundle was reduced and the self-tuning mechanism was replaced by a constant
    \ca flow from the cell body.}
  \label{fig_relaxational}
\end{figure}

The self-tuning mechanism described above is sufficiently robust to adjust
the hair bundle to the critical point despite the natural variation
of conditions that is likely to occur {\it
 in vivo}. Many experimental
 
preparations, however, involve significant perturbation from natural
conditions, and this may shift the system into a different dynamical
regime. In the following, we consider a particular type of dynamical behavior
that can arise when the calcium feedback is weak.  In this situation, as
previously discussed in Fig.~\ref{fig_fixed}a, the interaction of the
transduction channels with the calcium flux can produce a bistable system. If
the position $\ym$ of the adaptation motors is fixed, the hair bundle will
settle at one or other of the stable positions. However, this state of affairs
is upset by the slow calcium feedback acting on the motors
(Eq.~\ref{eq_mot}). Suppose, for example, that the bundle is at the fixed
point with the higher value of $x$, for which there is a high probability that
the transduction channels are open.  The \ca ions entering through the channel
bind to the motors, causing a fraction of them to detach, and the diminishing
force exerted by the motors causes the tension in the tip links $\ttl$ to
fall. As indicated in Fig.~\ref{fig_relaxational}a, the fixed point vanishes
at a critical value of $\ttl$, and the system then abruptly jumps to the other
fixed point. At this lower value of $x$, the channels are mostly closed.  The
resulting drop in calcium concentration augments the number of bound motors,
increasing the tension in the tip links until the lower fixed point becomes
unstable, whereupon the hair bundle jumps back to its initial position.

The consequence of this dynamics is a slow relaxation oscillation
 \citep[][Sect.~7.5]{Strogatz1994} of the bundle deflection $X$, which has a
non-sinusoidal form: each abrupt jump in position is followed by an interval of
slow motion, as shown in Fig.~\ref{fig_relaxational}b. Martin, Mehta and
Hudspeth (\citeyear{Martin.Hudspeth2000}) have observed oscillations of this
type in their preparations of dissected tissue. They attributed the dynamical
behavior to precisely the motor-mediated mechanism that we discuss here, and
speculated that these oscillations form the basis of mechanical
amplification of auditory stimuli. According to our model however, the
relaxation oscillations differ in many respects from the Hopf oscillations
which provide active amplification. They are considerably slower; their
frequency is strongly amplitude-dependent; and the frequency does not
significantly depend on the viscous relaxation time of the bundle, but is
principally governed by the motors' sensitivity to changing calcium levels. In
our model, relaxation oscillations are an accidental consequence of the
self-tuning mechanism when the system is too far from normal operating
conditions.

In the light of our analysis, various interpretations of the experiment of
Martin, Mehta and Hudspeth \citeyear{Martin.Hudspeth2000} are possible.  It
may be that in saccular hair cells, which are designed to detect low frequency
vibrations, slow motor-mediated relaxation oscillators are sufficient to
provide a rudimentary amplification of signals. Another possibility is that
the observed oscillations can be adjusted to the vicinity of a Hopf
bifurcation by some mechanism that we have not considered here, e.g. by
alternating the bundle stiffness. A third possibility to bear in mind is that
the conditions in preparations of dissected tissue are not precisely the same
as those {\it in vivo}, and thus the hair bundles may not be operating in the
dynamical regime that is physiologically relevant. Careful adjustment might
permit the channel adaptation and the motor-mediated self-tuning mechanism to
work together as we propose they do in auditory hair cells, in which case more
rapid, critical Hopf oscillations might be observed.

\section*{Conclusion}

Our phenomenological model illustrates that the two adaptation processes which
have been documented in hair cells can, together, generate a self-tuned Hopf
bifurcation. The feedback mechanism that accomplishes the self-tuning
requires, in addition, a system to detect bundle oscillations.  We have shown
that this can be accomplished by the resonant response of the BK and
voltage-gated \ca channels which has been established experimentally in lower
vertebrates. The spontaneous frequency $\omc$ of a hair cell is specified
principally by the architecture of its hair bundle. To ensure robust operation
of a hair cell, we expect that both the calcium relaxation rate $\lambda$, and
the resonant frequency of the electrical oscillator are quite closely matched
to the spontaneous oscillation frequency $\omc$.  The self-tuning mechanism
can reliably keep the hair bundle poised on the verge of oscillating {\it in
vivo}. But the perturbations in conditions which can occur in some
experimental preparations may lead to quite different dynamical regimes. Thus,
while recent experiments on hair cells of the bullfrog sacculus show promising
signs of active amplification
\citep{Martin.Hudspeth1999,Martin.Hudspeth2000,Martin.Julicher2001,Martin.Hudspeth2001},
they may well not yet have revealed the full potential of auditory hair cells
to detect faint sounds.

\section*{Appendix}
\subsection*{Series expansion}
In this section we derive an analytical expression for the third-order
coefficient $\B$ in the complex equation of motion (Eq.~\ref{eq_complex}).
The Taylor expansion of $\dot \z$ up to
third order, with the use of a new variable
$\xi=\Delta x-\Delta x_s(C)$ can be derived from the equations of motion
(Eqs.~\ref{eq_dotc} and \ref{eq_ydot}) and the definition of $\z$
(Eq.~\ref{eq_z}): 
\begin{multline}
\dot \z = \left[ \mbox{Linear terms in $\Delta x$, $\Delta C$} \right]  \\
- \frac{N \gamma^2  }{\zeta}\left[f' \xi +\frac 12 f'' \xi^2 + \frac 16 f'''
  \xi^3\right]  \\ + \alpha \lambda C_M \left[ \po' \xi + \frac12 \po'' \xi^2 + \frac 16 \po ''' \xi^3 \right]\;.
\end{multline}
Expanding $\xi$ as a Taylor series in $\Delta x$ and $\Delta C$, expressing
the latter in terms of $\z$ using Eq.~\ref{eq_solution} 
\begin{multline}
\xi=\frac{(\alpha +x_s')\bar \z -(\bar \alpha +x_s') \z}{\alpha - \bar \alpha}
- \frac12 x_s'' \frac{(\z -\bar \z)^2}{(\alpha - \bar \alpha)^2}\\
- \frac16 x_s'''  \frac{(\z -\bar \z)^3}{(\alpha - \bar \alpha)^3}+\mathcal{O}(\z^3)
\end{multline}
and collecting the real part of the terms containing $\z^2 \bar \z$ yields
\begin{multline}
  \re \B =
 - \frac{N \gamma^2 }{16\zeta (\im \alpha)^2}
  \left(-x_s'' f''
    +\left| \alpha+x_s' \right|^2 f''' \right)  \\
  -\frac{\lambda C_M}{16(\im \alpha)^2} \Bigl(
  x_s''' \po'
    -x_s''(2 \re \alpha+ 3 x_s')\po'' \\+x_s'\left| \alpha+x_s' \right|^2\po''' \Bigr) \;.
\end{multline}
Inserting $\alpha=x_s'(-\lambda (1+\mu) +\I \omc)/(\lambda \mu)$ from
 Eq.~\ref{eq_alpha} gives
\begin{multline}
  \re \B =
 -\frac{(\lambda \mu)^2}{16 (\omc x_s')^2}\Biggl( \frac{N \gamma^2 }{\zeta}
  \Biggl(-x_s'' f'' \\
    +(x_s')^2\left(1+\frac{\omc^2}{(\lambda\mu)^2}\right) f''' \Biggr)  
  +\lambda C_M \Biggl(
    x_s''' \po'
 \\   -x_s'' x_s' (1-2/\mu)\po'' +(x_s')^3  \left(1+\frac{\omc^2}{(\lambda\mu)^2}\right)
 \po''' \Biggr) \Biggr)
\end{multline}
and using the specific form of $\Delta x_s$, given by Eq.~\ref{eq_po}, we finally obtain
\begin{multline}
  \re \B =
 - \frac{(\lambda \mu)^2}{16 \omc^2}
\Biggl(  \frac{N \gamma^2 }{\zeta}
\left[ \frac{f''}{D} +\left(1+\frac{\omc^2}{(\lambda\mu)^2}\right) f'''
 \right]\\
+\frac{\lambda C_M}{C} \left[ \frac{2 \po'}{D} - (1-2/\mu) \po'' +D
 \left(1+\frac{\omc^2}{(\lambda\mu)^2}\right) \po''' \right] \Biggr) \;.
\end{multline}
If we assume a matching between the \ca relaxation rate and the characteristic
bundle frequency, $\lambda \sim \frac {\kb}{\zeta} \sim \omc$,
the position $x$ of the bifurcation becomes independent of the
characteristic frequency and so do the terms $\po$, $f$ and all their
derivatives, evaluated at the bifurcation. 
The third-order coefficient then scales as 
\begin{equation}
\re \B \sim -\frac{\omc}{d_{\rm O}^{2}}\;.
\label{eq_bscale}
\end{equation}
The parameters we use in our simulation give the value $\re \B = - 6.7\,
 \omc
/ d_{\rm O}^{2}$.

\subsection*{Noise}
In order to determine the level of the smallest stimulus that a hair bundle can detect,
stochastic fluctuations of the bundle must be taken into account. There are several 
sources of noise in the system.  The most important is the Brownian force that the
surrounding fluid exerts on the bundle.  The noise resulting from the stochastic opening
and closing of channels is less significant, as can be seen by considering
a simplified two-state channel model. We denote the lever arm shift between the two 
states by $d$, the average dwell times as
$\tau_0$ and $\tau_1$ and the probability of fining the channel in state 1 by
$p_1$. Then the expectation value of the number of channels in state 1 is
$\left< n_1 \right> = N p_1$, and the variance is $\left< \Delta n_1^2 \right>=N
p_1 (1-p_1)$.  The fluctuation in the total force from all channels is
$\left< \Delta F^2 \right>=(\ktl d)^2 \left< \Delta n_1^2 \right>$.  The noise
has the spectral density of a dichotomous process
\begin{equation}
S(\omega)=\frac{\zeta}{\gamma^2} k_B T_N  \frac{1}{\pi \left( 1 + (\omega
    \tau_C)^2 \right) }\;.
\end{equation}
It has the characteristic of a Brownian noise with an effective temperature
$k_B T_N= \frac{\gamma^2}{\zeta} (\ktl d)^2 N p_1 (1-p_1) \tau_C \approx k_B T
\omc \tau_C$, where $\tau_C=\left( \tau_0 ^{-1} +\tau_1 ^{-1}\right)^{-1}$.
 The second approximation is based on the estimates $k_B T \sim \ktl d_{\rm
  O}^2$ (from the condition for the negative stiffness) and $\omega_C\sim N
\gamma^2 \ktl / \zeta$ (Eq.~\ref{eq_scale_omega} with $\lambda \sim
\omega_C$). 
Since the channel switching time is shorter than the period of the
oscillations, $ \omc \tau_C \ll 1$, it follows that the channel noise is
negligible compared to the thermal noise.  The same can be shown for the
influence of stochastic channel opening on the \ca concentration inside the
stereocilia.

Taking the Brownian force exerted by the fluid into account, the equation of
motion for the complex variable $\z$ becomes
\begin{equation}
\dot \z = (\epsilon +\I \omc) \z + \B |\z|^2 \z + \F + 
\frac{\gamma^2}{\zeta} F_N (t)
\end{equation}
where $F_N(t)$ represents white noise with the autocorrelation function
$\left< F_N(t) F_N(t') \right>= [2 \zeta k_B T /\gamma^2 ]\delta(t-t')$.
Writing $\z'=\z e^{-\I \omega t}$ (and taking the stimulus frequency
to be equal to the characteristic frequency), this
equation can be transformed to
\begin{align}
\frac{\zeta}{\gamma^2} \dot \z' &= -\frac{\partial}{\partial \z'} U (\z') +F_N (t) e^{-\I
\omega t}\;, \\
U(\z')&=\frac{\zeta}{\gamma^2} \left(
  -\frac{\epsilon}{2} \left| \z' \right| ^2-\frac {\re \B} 4 \left| \z'
\right|^4 -|\F| \z' \right) \;.
\label{eq_langevin}
\end{align}
It can be further simplified if we replace the ``rotating'' noise $F_N (t)
e^{-\I \omega t}$ (which reflects the fact that the noise only acts on the
spatial coordinate and not on the \ca concentration) with an averaged,
homogeneous noise term.  The averaged noise then has the temperature $\frac 12
T$ and Eq.~\ref{eq_langevin} becomes a Langevin equation for a particle of
friction coefficient $\zeta/\gamma^2$, moving in a potential $U(\z')$. The
equilibrium solution is the Boltzmann distribution $P(\z') \propto \exp(-
U(\z')/\frac 12 k_B T)$.  For a bundle that is tuned to the critical point
($\epsilon=0$), the mean-square fluctuation amplitude in the absence of a
stimulus ($\F = 0$) is
\begin{equation}
\label{eq_fluctuations_bif}
\left< \z^2 \right> =\sqrt{\frac{2 k_B  T \gamma^2}{\pi \zeta (-\re \B)}}
\sim \frac{d_{\rm O}^2}{\sqrt{N}}
\end{equation} 
(in the second step we used the estimates $\re \B \sim -\omega/d_{\rm O}^2$
(Eq.~\ref{eq_bscale}) and $k_B T \sim \ktl d_{\rm O}^2$, the condition for
negative stiffness).
Using the numerical parameter values we obtain
\begin{equation}
\nonumber
\left< \z^2 \right> \approx 0.13\, \frac{d_{\rm O}^2}{\sqrt{N}}\;.
\end{equation}
Weak stimuli which evoke a response that is smaller than the stochastic
fluctuations of the bundle, nevertheless cause phase-locking of the spontaneous
bundle oscillations \citep{Camalet.Prost2000,Martin.Hudspeth2001}.  
The brain might detect such signals through periodicity in the
spike-train elicited by individual hair cells, or via the synchronicity of spikes from
neighboring hair cells, whose characteristic frequencies differ only
slightly. 
In either case, the quantity of interest is the Fourier component of bundle oscillations
at the sound frequency. With a finite stimulus force $\F$, the Fourier component is
\begin{equation}
\label{eq_fourier}
\left< \z e^{-\I \omega t} \right> =\sqrt{\frac{2}{\pi}} 
\frac{|\F|}{\sqrt{-\re \B k_B T \gamma^2/\zeta}} \sim
A \nu \sqrt{N}  \frac{\db}{L} \;.
\end{equation}

We assume that signals can be detected provided that the signal-to-noise ratio, 
$\left< \z e^{-\I \omega t} \right> / \sqrt{\left< \z^2 \right>}$, exceeds a certain 
threshold value $\rho$. The smallest detectable amplitude of fluid motion then scales 
as
\begin{equation}
  \label{eq_ampl_min}
   A_{\rm min} \sim \frac{\rho d_{\rm O}}{\nu} \frac {L}{\db} N^{-\frac{3}{4}}\;
\end{equation}
or with our numerical parameter values
\begin{equation}
  \nonumber
  A_{\rm min} \approx 1.0\, \frac{\rho d_{\rm O}}{\nu} \frac {L}{\db} N^{-\frac{3}{4}}\;.
\end{equation}
Taking $\rho \approx 0.2$, $L/b\approx 3$,  $\nu \approx 1$, and $N\approx 200$
gives $A_{\rm min} \approx 0.1\;{\rm nm}$ for high frequency bundles containing
many stereocilia. 

\begin{table*}
\caption{Parameter values, used in the simulation}
\label{tab:parameters}
\begin{center}
\begin{tabular}{llll}
\hline
Tip-link stiffness & $\ktl$ & $0.5\,{\rm pN/nm}$ &
\citep{Martin.Hudspeth2000}\\
Channel states & $d_{C2}$ & $7\,{\rm nm}$ \\
 & $d_{O}$ & $8.5\,{\rm nm}$ & \citep{Martin.Hudspeth2000}\\
Channel state energies & $G^0_{C2}$ & $60\,{\rm zJ}$\\
 & $G^0_{O}$ & $70\,{\rm zJ}$\\
\ca dependent gating point shift & $D$ & $4\,{\rm nm}$ & \citep{Corey.Hudspeth1983b} \\
Switching time & $\tau_{\rm ch}$ & $50\,\mu{\rm s}$ & \citep{Corey.Hudspeth1983b}\\
Distance between stereociliary roots & $b$ & $1\,\mu{\rm m}$ & \citep{Howard.Hudspeth1988a}\\
Bundle height & $L$ & $7\,\mu{\rm m}$ & \citep{Howard.Hudspeth1988a} \\
Geometric factor & $\gamma$ & $0.14$ & \citep{Howard.Hudspeth1988a} \\
Pivotal stiffness of a stereocilium &  $\kappa$ & $1.5 \times 10^{-16} \,{\rm
  Nm/rad}$ & \citep{Hudspeth.Martin2000}\\
Number of stereocilia & $N$ & $50$ & \citep{Martin.Hudspeth2000}\\
Viscous friction & $\zeta$ & $0.65\,{\rm pN\,ms/nm}$\\
\ca relaxation rate & $\lambda$ & $100\,{\rm s}^{-1}$\\
Channel contribution to the \ca concentration & $C_M$ & $7\,\mu {\rm M}$ &
\cite{Lumpkin.Hudspeth1998} \\
Number of adaptation motors per channel & $N_M$ & 20 \\
Motor association rate & $\omega_M$ & $200\,{\rm s}^{-1}$ \\
Motor speed & $v_M$ & $0.3\,\mu{\rm m/s}$\\
Motor stall force & $F_M$ & $1\,{\rm pN}$ \\
Maximum transduction current & $I_{\rm Input}/N$ & $500\,{\rm pA}$ &
\citep{van.Netten.Kros2000} \\
Filter resistance& $R_F$ & $35\,{\rm mV/nA}$\\
Filter Q-factor & $Q$ & $10$& \citep{Hudspeth.Lewis1988a}\\
Filter frequency & $\omega_F$ & $2\pi \times 50\,{\rm s}^{-1}$\\
Membrane potential & $V_0$ & $-70\,{\rm mV}$ \\
\ca channels gating point & $V_{\rm Ca}$ & $-65\,{\rm mV}$ \\
\ca channels gating charge & $q$ & $q/{k_BT}=0.2\,{\rm mV}^{-1}$ \\
Time constant of the slow feedback & $\lambda_{\rm B}$ & $1\,{\rm s}^{-1}$\\
\ca gain due to slow feedback & $I_{\mbox{\ca}}$ & $10\,\mu{\rm M/s}$ \\
\hline
\end{tabular}
\end{center}
\end{table*}

\subsection*{Simulation of self-tuned critical oscillations}

The model parameters and their values, used in the simulation, are listed in
Table \ref{tab:parameters}. 
The channels were simulated as described in the main text, using a three-state
model with parameters $G^0_{\rm C1}=0$, $G^0_{\rm C2}= 60 \,{\rm zJ}$,
$G^0_{\rm O}= 70 \,{\rm zJ}$,
$d_{\rm C1}=0\,{\rm nm}$, $d_{\rm C2}=7\,{\rm nm}$, $d_{\rm O}=8.5\,{\rm nm}$
and $D=4\,{\rm nm}$.  The transition rate between the states C1 and C2 and
between C2 and O was modeled as $r_{i\to j}=\tau_{\rm ch}^{-1}
\exp((G_i-G_j)/2k_B T)$, with the free energy in each state $G_i=\frac12 \ktl
(x-x_s(C)-d_i)^2 +G^0_i$.

The resonant filter was simulated using the differential equation corresponding
to the filter characteristics given by Eq.~\ref{eq_filter}
\begin{equation}
\ddot{V}=-\omega_F^2 (V-V_0)-\frac{\omega_F}{Q} \dot V + (n_{\rm O}/N-p_{\rm O}^0) I_{\rm  Input} R_{F} \omega_F^2
\end{equation} 
where $n_{\rm O}$ is the number of open transduction channels 
and the parameters $R_F=35\,{\rm mV/nA}$, $I_{\rm Input}=0.5\,{\rm nA}$,
$Q=10$, $V_0=-70\,{\rm mV}$, $p_{\rm O}^0=0.15$ and $\omega_F=2\pi\times 
50 \,{\rm
  s}^{-1}$.

The calcium contribution from the cell body was modeled according to
Eq.~(\ref{eq_ca_simulation})
with $I_{Ca}=10\,{\rm \mu M/s}$, $\lambda_{\rm B}=1\,{\rm s}^{-1}$,
$V_{Ca}=-65\,{\rm mV}$, $q/k_BT=0.2\,{\rm mV}^{-1}$. 

The adaptation motors were modeled with stochastic attachment at a fixed
rate $\omega_M=200\,{\rm s}^{-1}$ and detachment at a \ca-dependent rate 
$g(C) \omega_M$, with $g(C)=(0.1\,{\rm \mu M}^{-3})\times C^3$. 

All differential equations were solved using the Euler method. The time step
was $1\,{\mu s}$.  

\section*{Acknowledgment}
 We would like to thank R. Fettiplace, G. Manley, P. Martin, and S. van Netten
for helpful discussions and comments on the manuscript. A.V. is a European
Union Marie Curie Fellow.  T.D. is a Royal Society University Research Fellow.

\end{document}